\documentclass[twocolumn,twoside,pra,superscriptaddress]{revtex4}

\usepackage[pdftex]{graphicx,color}
\usepackage{amsmath}
\usepackage{amssymb}
\usepackage{epsfig}
\usepackage{amsthm}
\usepackage{tikz}
\usepackage{wrapfig}
\usepackage{comment}
\usepackage{latexsym,revsymb}

\newcommand{\ket}[1]{| #1 \rangle}
\newcommand{\bra}[1]{\langle #1 |}

\begin{document}
\title{Discrete Spacetime and Relativistic Quantum Particles}
\author{Terence C.\ Farrelly} \email{tcf24@cam.ac.uk} \affiliation{DAMTP, Centre for Mathematical Sciences, Wilberforce Road, Cambridge, CB3 0WA, United Kingdom}
\author{Anthony J.~Short}\email{tony.short@bristol.ac.uk}\affiliation{H. H. Wills Physics Laboratory, University of Bristol$\text{,}$ Tyndall Avenue, Bristol, BS8 1TL, United Kingdom}

\begin{abstract}
We study a single quantum particle in discrete spacetime evolving in a causal way.  We see that in the continuum limit any massless particle with a two dimensional internal degree of freedom obeys the Weyl equation, provided that we perform a simple relabeling of the coordinate axes or demand rotational symmetry in the continuum limit.  It  is surprising that this occurs regardless of the specific details of the evolution: it would be natural to assume that discrete evolutions giving rise to relativistic dynamics in the continuum limit would be very special cases.  We also see that the same is not true for particles with larger internal degrees of freedom, by looking at an example with a three dimensional internal degree of freedom that is not relativistic in the continuum limit.  In the process we give a formula for the Hamiltonian arising from the continuum limit of massless and massive particles in discrete spacetime.
\end{abstract}

\maketitle
\section{Introduction}
Approximating physical systems in continuous spacetime by discrete systems is an important challenge in physics.  For example, to simulate physics in the continuum one typically discretizes spacetime and other degrees of freedom.  Also, it is often useful to define quantum field theories in continuous spacetime as the continuum limit of quantum field theories in discrete spacetime \cite{Creutz83}.  Furthermore, it is tempting to speculate that spacetime might be discrete at some small scale.  A prominent example of this is causal set theory \cite{BLMS87}.  Whatever the motivation, if discrete spacetime models are to be useful, they must approximate the dynamics of continuous physical systems at low energies.

Here we will study a single quantum particle evolving in a causal and translationally invariant way in discrete spacetime, where causal means that there is a maximum speed of propagation of information.  In fact, for this to be possible in discrete space, we {\it must} take time to be discrete \cite{FS13}.  Furthermore, in order to obtain non-trivial dynamics we must give the particle an internal `spin' degree of freedom \cite{Meyer96a}.  Such single particle evolutions are examples of discrete-time quantum walks, which are useful in quantum computing \cite{Kempe03}.

To show that such discrete dynamics approximate physical systems in continuous spacetime, we take the continuum limit of the discrete evolution.  Our main result is that the continuum limit of the evolution
of a discrete massless particle with an additional two dimensional degree of freedom is always equivalent to a particle obeying the relativistic Weyl equation ($\textrm{d}\psi(t)/\textrm{d}t=\pm\vec{\sigma}.\vec{P}\psi(t)$) if we relabel the coordinate axes in a simple way (by rotating, rescaling and removing a constant  velocity shift).  Alternatively, if such a discrete evolution is chosen to have rotational symmetry in the continuum limit, then it must obey the Weyl equation.  Discrete models cannot have continuous spacetime symmetries, so it is surprising that the emergence of Lorentz symmetry is generic for these models.

The Weyl and Dirac equation describe the free evolution of spin-half fermions in the continuum, which, together with bosonic fields, are the basic constituents of nature.  Finding discrete causal models that may reproduce these systems in the continuum limit is a useful endeavour, particularly because such models may be well suited to simulation by quantum computers \cite{FS13,D'Ariano12b}.

Examples of quantum particles in discrete spacetime have been studied in connection with relativistic dynamics in \cite{FH65,Bial94,Meyer96,Meyer97,Strauch06,BES07,Strauch07,Kurz08}. In particular, \cite{Bial94} has examples of discrete quantum particles that obey the three dimensional Weyl and Dirac equations in the continuum limit.  In fact, by making some requirements on how the evolutions transform under rotations, \cite{Bial94,DP13} show that discrete evolutions with a body centred cubic neighbourhood and two dimensional extra degrees of freedom obey the Weyl equation in the continuum limit.

After introducing notation, we discuss causal quantum particles in discrete spacetime in section \ref{sec:Properties of Quantum Particles in Discrete Spacetime}.  Then in section \ref{sec:Taking the Continuum Limit} we take their continuum limit.  In section \ref{sec:The Continuum Hamiltonian} we show that, if the discrete particle is massless and has a two dimensional extra degree of freedom, then it obeys the Weyl equation in the continuum limit.  In section \ref{sec:Reproducing the Weyl equation in three space dimensions} we reproduce the discrete evolution given in \cite{Bial94} that becomes a particle obeying the Weyl equation in the continuum limit.  In section \ref{sec:More than two extra degrees of freedom} we see that the continuum limit of massless systems with more than two extra degrees of freedom may have rotational but not necessarily Lorentz symmetry.  In section \ref{sec:Mass and the Dirac Equation} we look at the continuum dynamics with mass included.  We conclude with a discussion in section \ref{sec:discussion}.

\section{Setup}
We label discrete space coordinates by vectors $\vec{n}$, where each of the $d$ components of $\vec{n}$ takes integer values.  Then the orthornormal basis $\ket{\vec{n}}$ of the Hilbert space $\mathcal{H}_P$ describes the particle's position.  The particle also has a finite dimensional extra degree of freedom described by states in $\mathcal{H}_S$, so its total state space is $\mathcal{H}_P\otimes\mathcal{H}_S$.  The extra degree of freedom will often correspond to spin or chirality in the continuum limit.

We are assuming time translation invariance, so the evolution operator $U_D$ is the same for every timestep.

We denote the identity on a Hilbert space $\mathcal{H}_X$ by $\openone_X$.  And, if, for example, $A$ is an operator on $\mathcal{H}_{Y}$ and $\psi$ is a vector in $\mathcal{H}_{X}\otimes\mathcal{H}_{Y}$, then we write $A\psi$ to mean $(\openone_{X}\otimes A)\psi$.

We will mostly be interested in particles that obey the Weyl equation in the continuum limit, which is the equation of motion of massless chiral fermions.  This means that they evolve via the Weyl Hamiltonian $H=\pm\vec{\sigma}.\vec{P}$, with $c=\hbar=1$.  The components of $\vec{\sigma}$ are the three Pauli operators, which act on the particle's spin, and $\vec{P}$ is the momentum operator.  The plus sign correpsonds to right handed particles and the minus sign corresponds to left handed particles \footnote{We can rewrite the right and left handed Weyl equations in a form that makes their Lorentz invariance more obvious: $i\sigma^{\mu}\partial_{\mu}\psi(x)=0$ and $i\overline{\sigma}^{\mu}\partial_{\mu}\psi(x)=0$ are the right and left handed Weyl equations respectively, where $\sigma^{\mu}=(\openone,\vec{\sigma})$ and $\overline{\sigma}^{\mu}=(\openone,-\vec{\sigma})$.  Lorentz invariance follows because $\sigma^{\mu}$ and $\overline{\sigma}^{\mu}$ transform like four vectors under Lorentz transformations.}.

\section{Properties of Quantum Particles in Discrete Spacetime}
\label{sec:Properties of Quantum Particles in Discrete Spacetime}
To get some intuition, it is useful to look at a simple example.  Suppose we have a particle on a discrete line of points, with an extra degree of freedom described by the orthonormal states $\ket{r}$ and $\ket{l}$.  One possible evolution is
\begin{equation}
U_D=S\ket{r}\bra{r}+S^{\dagger}\ket{l}\bra{l},
\end{equation}
where $S$ is the unitary shift operator that takes the position state $\ket{n}$ to $\ket{n+1}$.  But this evolution is not terribly interesting: $U_D$ merely shifts all $\ket{l}$ states to the left and all $\ket{r}$ states to the right.  Instead, we can consider the new evolution
\begin{equation}
 U_D=W\left(S\ket{r}\bra{r}+S^{\dagger}\ket{l}\bra{l}\right),
\end{equation}
where $W$ is  a unitary operator  on $\mathcal{H}_S$.

With initial state $\ket{r}\ket{0}$, $U_D$ first shifts the position from $\ket{0}$ to $\ket{1}$, and then $W$ takes $\ket{r}$ to a superposition of $\ket{r}$ and $\ket{l}$.  Over the next timestep, because the state now has overlap with both $\ket{l}$ and $\ket{r}$, the particle spreads out and is effectively slowed down.  This is a simple discrete analogue of how mass mixes chiralities in the Dirac equation.

Let us now consider a general causal quantum particle on a lattice.  Translational invariance allows us to write the evolution operator in a simple form. First,
\begin{equation}
 U_D=
\displaystyle\sum_{\vec{n},
\vec{q}}A^{\vec{n}}_{\vec{q}}\ket{\vec{n}+\vec{q}}\bra{\vec{n}},
\end{equation}
where $A^{\vec{n}}_{\vec{q}}=\bra{\vec{n}+\vec{q}}U_D\ket{\vec{n}}$ is an operator on $\mathcal{H}_S$.  Translational invariance means $A^{\vec{n}}_{\vec{q}}$ does not depend on $\vec{n}$.  With $A_{\vec{q}}=A^{\vec{n}}_{\vec{q}}$, and defining $S_{\vec{q}}$ to be the operator that shifts a position state by $\vec{q}$, we have
\begin{equation}
 U_D=\displaystyle\sum_{\vec{q}}A_{\vec{q}}S_{\vec{q}}.
\end{equation}
We also impose causality, so that $A_{\vec{q}}$ will only be non-zero for some finite set of vectors $\vec{q}$.

Note that an extra degree of freedom is required for these particles to have non trivial evolution, where trivial means $U_D$ is just proportional to a shift operator \cite{Meyer96a}.

Finally, before we take the continuum limit, we will define massive and massless evolution.  Unitarity implies that
\begin{equation}
 U_D^{\dagger}U_D=\displaystyle\sum_{\vec{q}}A^{\dagger}_{\vec{q}}S^{\dagger}_{\vec{q}}\displaystyle\sum_{\vec{p}}A_{\vec{p}}S_{\vec{p}}=\openone_D.
\end{equation}
But terms like $S^{\dagger}_{\vec{q}}S_{\vec{p}}$ with $\vec{q}\neq\vec{p}$ must vanish, so it follows that
\begin{equation}
 \displaystyle\sum_{\vec{q}\neq\vec{p}}A^{\dagger}_{\vec{q}}A_{\vec{p}}=0\ \textrm{and}\ 
 \displaystyle\sum_{\vec{q}}A^{\dagger}_{\vec{q}}A_{\vec{q}}=\openone_S,
\end{equation}
which implies that $\sum_{\vec{q}}A_{\vec{q}}$ is a unitary operator on $\mathcal{H}_S$.  This allows us to write
\begin{equation}
\label{eq:2}
 U_D=W\displaystyle\sum_{\vec{q}}A^{\prime}_{\vec{q}}S_{\vec{q}},
\end{equation}
where $W=\sum_{\vec{q}}A_{\vec{q}}$ is a unitary on $\mathcal{H}_S$ and $A^{\prime}_{\vec{q}} = W^{\dagger} A_{\vec{q}}$ such that $\sum_{\vec{q}}A^{\prime}_{\vec{q}}=\openone_S$.  Then, analogously to the example at the beginning of this section, if $W=\openone_S$, we say that the particle is massless.

For now we will focus on massless evolutions, but later in section \ref{sec:Mass and the Dirac Equation} we will look at continuum limits of massive evolutions.  In the massive case, one way to ensure that the dynamics will have a continuum limit is to let $W$ tend to $\openone_S$ as the length of the timestep, $\delta t$, goes to zero. 

In a sense, massless evolutions seem more natural because to take the continuum limit we need only shrink the lattice spacing and the length of the timestep; the evolution on the lattice remains the same.  On the other hand, for massive evolutions we need to make the discrete evolution dependent on the lattice scale to get a continuum limit \footnote{Although this is necessary to get the Dirac equation as the continuum limit of a discrete evolution (see section \ref{sec:Mass and the Dirac Equation}), it is reassuring to note that in the standard model fermions are fundamentally massless and only acquire mass through the Higgs mechanism.}.

\section{Taking the Continuum Limit}
\label{sec:Taking the Continuum Limit}
Now we will take the continuum limit of these discrete evolutions.  The discrete evolution operator is
\begin{equation}
 U_D=\displaystyle\sum_{\vec{q}}A_{\vec{q}}S_{\vec{q}},
\end{equation}
which has the corresponding continuum Hamiltonian
\begin{equation}
  H=\displaystyle \left( \frac{a}{\delta t} \right) \sum_{\vec{q}}A_{\vec{q}}(\vec{q}.\vec{P}),
\end{equation}
where  $a$ is the lattice spacing and $\delta t $ is the discrete time-step.
To see this, we look at states that are smooth over many lattice sites, which is equivalent to looking at the subspace of states with low momentum.

Discrete momentum states are
\begin{equation}
 \ket{\vec{p}}=\frac{1}{a^{d/2}}\displaystyle\sum_{\vec{n}}e^{i\vec{p}.\vec{n}a}\ket{\vec{n}},
\end{equation}
where the components of $\vec{p}$ take values in $(-\textstyle{\frac{\pi}{a}},\textstyle{\frac{\pi}{a}}]$.

Continuum momentum states are
\begin{equation}
 \ket{\vec{p}}=\int_{-\infty}^{\infty}\!\textrm{d}^dx\, e^{i\vec{p}.\vec{x}}\ket{\vec{x}},
\end{equation}
where the components of $\vec{p}$ take values in  $\mathbb{R}$.

Now we identify the discrete particle's momentum states with those of a continuum particle with the same value of $\vec{p}$.  When acting on states with high momentum the continuum and discrete evolutions will be very different.  But the two evolutions will be similar if we restrict to low momentum states.  Let us define $\mathcal{H}_{\Lambda}$ as the space spanned by states with $|\vec{p}|\leq \Lambda \ll \frac{\pi}{a}$, and define $\tilde{U}_D$ and $\tilde{H}$ to be the restriction of $U_D$ and $H$ to $\mathcal{H}_{\Lambda}$.

Consider a discrete evolution for $n$ time-steps of length $\delta t$, corresponding to a total evolution time $t=n \delta t$. 

To compare the discrete and continuum evolution, with the latter given by $e^{-iHt}$, on the low momentum subspace, we evaluate
\begin{equation}
 \|e^{-iHt}\ket{\psi_{\Lambda}}-U_D^{n}\ket{\psi_{\Lambda}}\|_2 \leq \|e^{-i\tilde{H}t}-\tilde{U}_D^{n}\|,
\end{equation}
where $\ket{\psi_{\Lambda}}\in\mathcal{H}_{\Lambda}$ and $\|\cdot\|$ is the operator norm on $\mathcal{H}_{\Lambda}$.  Next we use the inequality for unitaries, $U$ and $V$, $\|U^n-V^n\|\leq n\|U-V\|$ \cite{NC00a}.  It follows that
\begin{equation} 
 \|e^{-i\tilde{H}t}-\tilde{U}_D^n\|\leq n\|e^{-i\tilde{H}\delta t}-\tilde{U}_D\|.
\end{equation}
To bound the right hand side, note that the evolution operator for a discrete particle can be written as
\begin{equation}
  U_D=\displaystyle\sum_{\vec{q}}A_{\vec{q}}S_{\vec{q}}\equiv
 \displaystyle\sum_{\vec{q}}A_{\vec{q}}\exp(-i(\vec{q}.\vec{P})a),
\end{equation}
where $\vec{P}$ is the momentum operator.  By taking the Taylor expansions of both $e^{-i\tilde{H}\delta t}$ and $\tilde{U}_D$, we show in appendix \ref{sec:Bounding the Norm} that for sufficiently small values of $\Lambda a $
\begin{equation} \label{eq:otherbound}
 \|e^{-i\tilde{H}\delta t}-\tilde{U}_D\|\leq C(\Lambda a)^2,
\end{equation}
where $C$ is a constant.  (The bound for the massive case is slightly different.  See section \ref{sec:Mass and the Dirac Equation} for details.)  Then
\begin{equation} \label{eq:bound}
 \|e^{-iHt}\ket{\psi_{\Lambda}}-U_D^{n}\ket{\psi_{\Lambda}}\|_2\leq Ct\Lambda^2 \frac{a^2}{\delta t}.
\end{equation}

To get a continuum limit, we fix $t$ and let $a, \delta t \rightarrow 0$ in such a way that $\textstyle{a/\delta t}$ is constant.  Because $t$ is fixed, the number of timesteps $n$ must tend to infinity.  We also take $\Lambda \rightarrow \infty$ at a slower rate than $a \rightarrow 0$, such that $\Lambda^2 a \rightarrow 0$. As the right-hand side of (\ref{eq:bound}) tends to zero and the momentum cut-off tends to infinity, this tells us that the discrete evolution defined by $U_D$ converges to the continuum evolution generated by the Hamiltonian $H$.

\section{The Continuum Hamiltonian}
\label{sec:The Continuum Hamiltonian}
In this section we will look at the continuum Hamiltonian.  For now we will suppose that these particles live in three spatial dimensions.  At the end of the section we will comment on what changes when $d\neq 3$.

First we will see that, if we can construct a massless evolution with a two dimensional extra degree of freedom that has the rotational symmetries of the lattice in the continuum limit, it must also have Lorentz symmetry.

Suppose that the continuum Hamiltonian has the rotational symmetries of the lattice.  The Hamiltonian is
\begin{equation}
\label{eq:1}
 H=\vec{B}.\vec{P},
\end{equation}
where $\vec{B}=\left( \frac{a}{\delta t} \right)\sum_{\vec{q}}A_{\vec{q}}\,\vec{q}$.  As each $B_i$ is Hermitian, we have $B_i=c_i\openone_S+\vec{n}_i.\vec{\sigma}$, with $c_i$ and $\vec{n}_i$ real.  That the evolution has the rotational symmetries of the lattice implies that there is a subgroup $G$ of $SU(2)$ whose action on $\{B_i:i=1,2,3\}$ is a representation of these symmetries.  Now, for a three dimensional lattice and a given $i$ and $j \neq i$ there must be a $V\in G$ such that $VB_iV^{\dagger}=-B_i$ and $VB_jV^{\dagger}=B_j$.  This implies that $c_i=0$, and also that $\textrm{tr}[B_i^{\dagger}B_j]=0$, which in turn means that $\vec{n_i}.\vec{\sigma}$ form an orthogonal set.  Furthermore, for any $i$ and $j$ there must exist a $V\in G$ such that $VB_iV^{\dagger}=B_j$, so we must have $|\vec{n_i}|=|\vec{n_j}|$.  It follows that $B_i$ are proportional to a representation of $\sigma_i$ or $-\sigma_i$.  We can modify the constant of proportionality by rescaling $a$ or $\delta t$. If we embed the lattice in the continuum with $a/\delta t$ chosen such that the constant of proportionality is one, the Hamiltonian will be equal to either the left or right handed Weyl Hamiltonian, which describes a Lorentz invariant evolution.

Now we will show that requiring rotational symmetry of $H$ is not quite necessary, meaning any massless discrete particle obeys the Weyl equation in the continuum limit if it has a two dimensional extra degree of freedom.

We can rewrite the Hamiltonian (in equation \ref{eq:1}) as
\begin{equation}
 H=\sigma_1\tilde{P}_{1}+\sigma_2\tilde{P}_{2}+\sigma_3 \tilde{P}_{3}+\vec{\beta}.\vec{P},
\end{equation}
	where $\vec{\beta}$ is a real vector and $\tilde{P}_{i}$ are real linear combinations of components of the momentum vector operator $\vec{P}$.  Now, the overall shift term $\vec{\beta}.\vec{P}$ is physically meaningless, so we remove it by changing to coordinates that are moving with a constant velocity $\vec{\beta}$.  This gets us closer to the Weyl Hamiltonian, but $\tilde{P}_i$ are not necessarily momentum operators in orthogonal directions.  To fix this we should think of 
$\sigma_1 \tilde{P}_{1}+\sigma_2 \tilde{P}_{2}+\sigma_3 \tilde{P}_{3}$ 
as a sum of tensor products of vectors since $\sigma_i$ and $\tilde{P}_j$ both span vector spaces.  Now we use the singular value decomposition (chapter $7$ of \cite{Horn85}) to rewrite $H$ as
\begin{equation*}
 H=\gamma_1\sigma_{1}^{\prime}P_{1}^{\prime}+\gamma_2\sigma_{2}^{\prime}P_{2}^{\prime}+\gamma_3\sigma_{3}^{\prime}P_{3}^{\prime},
\end{equation*}
where $\sigma_{i}^{\prime}$ are spin operators along orthogonal axes, $P_{i}^{\prime}$ are momentum operators along orthogonal spatial axes, and $\gamma_i$ are real numbers.  Note that we can choose $\sigma_i^{\prime}$ and $P_i^{\prime}$ to be real combinations of $\sigma_{i}$ and $P_j$ respectively \cite{Horn85}.  This is necessary so that $P^{\prime}_i$ and $\sigma_i^{\prime}$ have the right physical interpretation.  If all the $\gamma_i$ are non-zero, we can rescale the spatial axes so that $\gamma_i P^{\prime}_i\to P^{\prime}_i$.  Then, dropping primes, we get
\begin{equation}
H=\sigma_{1}P_{1}+\sigma_{2}P_{2}+\sigma_{3}P_{3}\equiv\vec{\sigma}.\vec{P},
\end{equation}
where $\sigma_i$ are a representation of the Pauli operators \footnote{We do not get a representation of $-\sigma_i$ because we may have done a reflection when going from $P_i$ to $P_i^{\prime}$.}.  If any of the $\gamma_i=0$, then the Hamiltonian is that of a lower dimensional Weyl equation.  This means that all massless discrete quantum particles with a two dimensional extra degree of freedom obey the Weyl equation in the continuum limit.  In the next section we reproduce an example of a discrete evolution that has this property.

In the argument above we had to relabel the coordinate axes to get the right answer.  Only if we had different particles with evolutions whose continuum limits could not be made into the same form by the {\it same} relabelling of the coordinate axes would there be any physical significance to the different forms of evolution in the continuum limit.

If the number of spatial dimensions is fewer than three, the same results apply but the particle obeys a lower dimensional Weyl equation.  If the number of spatial dimensions is greater than three, the particle still obeys the Weyl equation in at most three dimensions, meaning it does not move in the remaining directions.

\subsection{Reproducing the Weyl equation in three space dimensions}
\label{sec:Reproducing the Weyl equation in three space dimensions}
A discrete evolution in three dimensional space that becomes a Weyl particle in the continuum limit was first presented in \cite{Bial94}.  It works by preforming conditional shifts in each direction:
\begin{equation}
U_D=T_xT_yT_z,
\end{equation}
with
\begin{equation}
\label{eq:z}
 T_{b}=S_{b}\ket{\uparrow_{b}}\bra{\uparrow_{b}\!}+S^{\dagger}_{b}\ket{\downarrow_{b}}\bra{\downarrow_{b}\!}
\end{equation}
where $b\in\{x,y,z\}$, $S_{b}$ shifts one lattice site in the $b$ direction and $\ket{\uparrow_{b}}$ and $\ket{\downarrow_{b}}$ are spin up and spin down along the $b$ axis.  So, for example, $T_{z}$ shifts a particle in the state $\ket{\vec{n}}\ket{\uparrow_{z}}$ one step in the $+\hat{z}$ direction.

It is interesting that this discrete evolution essentially uses a body centred cubic neighbourhood.  In fact, the most obvious choice, the cubic neighbourhood, cannot give the three dimensional Weyl equation in the continuum limit \cite{Bial94}.

\subsection{More than two extra degrees of freedom}
\label{sec:More than two extra degrees of freedom}
Unfortunately, it is not true that discrete evolutions with more than two extra degrees of freedom become relativistic evolutions in the continuum limit.  Below is a simple example with a three dimensional extra degree of freedom, with basis states $\ket{1}$, $\ket{2}$ and $\ket{3}$.  In the continuum limit it becomes a single particle evolving via the Hamiltonian
\begin{equation}
 H=\vec{J}.\vec{P},
\end{equation}
where $J_i=-i \sum_{jk} \varepsilon_{ijk}\ket{j}\bra{k}$ are a three dimensional representation of the generators of the lie algebra of SO(3) acting on $\mathcal{H}_S$.  Although this has rotational symmetry, it does not have Lorentz symmetry \footnote{Note that we cannot add a term like $\vec{\beta}.\vec{P}$ to $H$ as we did in section \ref{sec:The Continuum Hamiltonian} because this would break rotational symmetry, as would rescaling coordinate axes.}.  To see this, note that $H^2-P^2$ is not Lorentz invariant \footnote{To see this, look at $\sum_{i}\bra{i}H^2-\vec{P}^2\ket{i}=-\vec{P}^2$.  If we are to have Lorentz invariance, $\sum_{i}\bra{i}U_{\Lambda}(H^2-\vec{P}^2)U^{\dagger}_{\Lambda}\ket{i}$ should be independent of the boost operator $U_{\Lambda}$.As we are talking about free particles, the effect of a Lorentz transformation is
$ U_{\Lambda}\ket{\vec{p}}\ket{k}=\sqrt{E_{\Lambda\vec{p}}/E_{\vec{p}}}\ket{\vec{\Lambda p}}D(\Lambda,\vec{p})\ket{k}$
where $D(\Lambda,\vec{p})$ is a unitary on the extra degree of freedom \cite{Weinberg95}.  But it follows from this that $\sum_{i}\bra{i}U_{\Lambda}(H^2-\vec{P}^2)U^{\dagger}_{\Lambda}\ket{i}=U_{\Lambda}(-\vec{P}^2)U^{\dagger}_{\Lambda}$, which is not independent of $\Lambda$.}.

The discrete evolution is a product of conditional shifts in each spatial direction:
\begin{equation}
U_D=T_xT_yT_z,
\end{equation}
but now with
\begin{equation}
T_b=\exp(-iaP_bJ_b),
\end{equation}
where $P_b$ is the momentum operator in the $b$ direction, with $b\in\{x,y,z\}$.  Also, we have relabelled $J_i$ by $x$, $y$ and $z$ in the usual way: $J_1=J_x$, $J_2=J_y$, and $J_3=J_z$.

To see the analogy with equation \ref{eq:z}, we can rewrite $T_b$ as
\begin{equation}
 T_{b}=S_{b}\ket{\!+\!1_b}\bra{+1_b}+\ket{0_b}\bra{0_b}+S^{\dagger}_{b}\ket{\!-\!1_b}\bra{-1_b},
\end{equation}
where $\ket{\lambda_b}$ is the eigenvector of $J_b$ with eigenvalue $\lambda$ and $S_{b}$ is a shift by one lattice site in the $b$ direction.

\section{Mass and the Dirac Equation}
\label{sec:Mass and the Dirac Equation}
Now we turn to evolutions with mass.  Recall that the evolution operator can be written as
\begin{equation}
 U_D=W\displaystyle\sum_{\vec{q}}A^{\prime}_{\vec{q}}S_{\vec{q}},
\end{equation}
where $W$ is a unitary on $\mathcal{H}_S$ and $\textstyle\sum_{\vec{q}}A^{\prime}_{\vec{q}}=\openone_S$.  To get a continuum limit, we will let $W$ tend to $\openone_S$ as $\delta t\rightarrow 0$ in the following way,
\begin{equation}
\label{eq:3}
 W=e^{-iM\delta t},
\end{equation}
with $M$ a fixed self-adjoint operator on $\mathcal{H}_S$.

The resulting continuum Hamiltonian is
\begin{equation}
 H=\left( \frac{a}{\delta t} \right)\displaystyle\sum_{\vec{q}}A_{\vec{q}}^{\prime}\,(\vec{q}.\vec{P})+M.
\end{equation}
To see this, we proceed exactly as in section \ref{sec:Taking the Continuum Limit}, with the only difference being a different upper bound for $\|e^{-i\tilde{H}\delta t}-\tilde{U}_D\|$, which is derived in appendix \ref{sec:Bounding the Norm with Mass}.  As in section \ref{sec:Taking the Continuum Limit} we let $a,\delta t\rightarrow 0$ to see that the discrete evolution agrees with the Hamiltonian above in the continuum limit.

As in the massless case, we can relabel coordinates so that the Hamiltonian becomes
\begin{equation}
 H=\vec{\sigma}.\vec{P}+M.
\end{equation}

In one space dimension, taking $M=m\sigma_x$, we get the Dirac Hamiltonian in one dimension:
\begin{equation}
 H=\sigma_z P_z+m\sigma_x.
\end{equation}

This is not generic, however.  For example, the choice $M=m_1\sigma_z+m_2\sigma_x$ is not a Lorentz invariant evolution \cite{Kurz08}.  That said, had we required emergent symmetry under a parity transformation, this Hamiltonian would not be allowed.

A discrete evolution that becomes a particle evolving via the Dirac equation in three spatial dimensions is given in \cite{Bial94}.  This works by taking two evolutions that give the left and right handed Weyl equations in the continuum limit and then mixing between them with a mass term.

\section{Discussion} \label{sec:discussion} 
We looked at the continuum limit of the evolution of a causal quantum particle in discrete spacetime.  In the massless case, when the particle had a two dimensional extra degree of freedom, we saw that the continuum limit evolution was essentially equivalent to that of a Weyl particle in three or fewer dimensions.  That such relativistic evolutions emerge generally in the continuum limit from discrete systems is exciting: it would have been reasonable to assume that discrete evolutions that are relativistic in the continuum limit would be very special cases.

These results for single particles naturally apply to free fermion fields in discrete spacetime evolving in a causal way.  The main challenge for the future is to find physically relevant {\it interacting} field theories evolving causally in discrete spacetime that have a continuum limit.  (One example that becomes the Thirring model in one spatial dimension is given in \cite{DdV87}.)

The evolutions we examined are discrete-time quantum walks, which first arose in quantum computation.  Also, causal (and potentially interacting) quantum systems in discrete spacetime can be viewed as Quantum Cellular Automata (a type of quantum computer) \cite{FS13,D'Ariano12a,D'Ariano12b,BDT12}.  So it is interesting to consider that applying ideas from quantum computation may help to understand the continuum limits of discrete quantum field theories \cite{JLP12}.

\acknowledgments
AJS acknowledges support from the Royal Society.  TCF acknowledges support from the Robert Gardiner Memorial Scholarship, CQIF, DAMTP, EPSRC and St John's College, Cambridge.

\bibliographystyle{unsrt}
\bibliography{QCA1,Other1,QFT1,Sim1,Walks1,Therm}

\appendix
\section{Bounding the Norm}
\label{sec:Bounding the Norm}
Here we bound $\|e^{-i\tilde{H}\delta t}-\tilde{U}_D\|$ for a massless evolution.  After Taylor expanding both terms, $\|e^{-i \tilde{H} \delta t}-\tilde{U}_D\|$ becomes
\begin{align}
 & \|\displaystyle\sum_{m\geq 2}\frac{(-i \tilde{H} \delta t)^m}{m!}-\displaystyle\sum_{\vec{q}}A_{\vec{q}}\displaystyle\sum_{l\geq 2}\frac{(-i\vec{q}.\vec{P}a)^l}{l!}\|\\
&\leq \displaystyle\sum_{m\geq 2}\frac{1}{m!}\|(-i \tilde{H} \delta t)^m-\displaystyle\sum_{\vec{q}}A_{\vec{q}}(-i\vec{q}.\vec{P}a)^m\|\\
&\leq \displaystyle\sum_{m\geq 2}\frac{a^m}{m!}\left(\|\displaystyle\sum_{\vec{q}}A_{\vec{q}}\,\vec{q}.\vec{P}\|^m+\|\displaystyle\sum_{\vec{q}}A_{\vec{q}}(\vec{q}.\vec{P})^m \|\right)\\
&\leq \displaystyle\sum_{m\geq 2}\frac{a^m}{m!}\left((\displaystyle\sum_{\vec{q}}\|A_{\vec{q}}\|\|\vec{q}.\vec{P}\|)^m+\displaystyle\sum_{\vec{q}}\|A_{\vec{q}}\|\|\vec{q}.\vec{P}\|^m\right)\\
&\leq \displaystyle\sum_{m\geq 2}\frac{a^m}{m!}\left((Kq\Lambda)^m+K(q\Lambda)^m\right)\\
&\leq 2 \displaystyle\sum_{m\geq 2}\frac{(Kq\Lambda a)^m}{m!} \\ &\leq C(\Lambda a)^2, \label{eq:l}
\end{align}
where $K$ is the number of $A_{\vec{q}}\neq 0$, $q$ is the largest value of $|\vec{q}|$ for which $A_{\vec{q}}\neq 0$ and the fifth line follows from $\|A_{\vec{q}}\|\leq 1$, which itself follows from $\sum_{\vec{q}} A_{\vec{q}}^{\dagger}A_{\vec{q}}=\openone_{S}$.  The last line applies when $\Lambda a \leq \frac{1}{K q }$ and follows from the fact that, when $\alpha \leq 1$, $\sum_{m\geq 2} \frac{\alpha^m}{m!} \leq \alpha^2\sum_{m\geq 2} \frac{1}{m!} =(e-2) \alpha^2=C^{\prime}\alpha^2$.

\section{Bounding the Norm with Mass}
\label{sec:Bounding the Norm with Mass}
Here we bound $\|e^{-i\tilde{H}\delta t}-\tilde{U}_D\|$ for a massive evolution.  We omit tildes now to simplify notation.  Define $U^{\prime}_D=W^{-1}U_D$, which is a massless discrete evolution with corresponding continuum Hamiltonian $H^{\prime}=H-M$. It follows from the triangle inequality that
\begin{equation}
\begin{split}
\|e^{-iH\delta t}-U_D\| \leq & \|e^{-iH\delta t}-e^{-iM\delta t}e^{-iH^{\prime}\delta t}\|\\
+ & \|e^{-iM\delta t}e^{-iH^{\prime}\delta t}-e^{-iM\delta t}U^{\prime}_D\|.
\end{split}
\end{equation}
The second term is $\|e^{-iH^{\prime}\delta t}-U^{\prime}_D\|$ because the operator norm is unitarily invariant.  We bounded this expression from above by $C(\Lambda a)^2$ in the previous section, so it remains to bound the first term.  To do this, note that the order one and order $\delta t$ terms cancel.  Then, by expanding in power series and using the triangle inequality, it follows that for sufficiently small $a$ (and hence $\delta t$)
\begin{equation}
\begin{split}
\|e^{-iH\delta t}-e^{-iM\delta t}e^{-iH^{\prime}\delta t}\|& \\
\leq C_1(\Lambda & a)^2 + C_2 \Lambda a\delta t + C_3\delta t^2,
\end{split}
\end{equation}
where $C_{i}$ are constants and $\Lambda$ is the momentum cutoff.  It follows that
\begin{equation}
   \|e^{-iH\delta t}-U_D\|
 \leq (C+C_1)(\Lambda a)^2 + C_2 \Lambda a\delta t + C_3\delta t^2.
\end{equation}
And so $\|e^{-iHt}-U_D^n\|\leq n\|e^{-iH\delta t}-U_D\|\to 0$ as $a$ tends to zero, provided we choose the momentum cutoff to grow sufficiently slowly with $a$.
\vfill
\end{document}